\documentclass[graybox]{svmult}

\usepackage{mathptmx}       
\usepackage{helvet}         
\usepackage{courier}        
\usepackage{type1cm}        
\usepackage{makeidx}         
\usepackage{graphicx}        
\usepackage{multicol}        
\usepackage[bottom]{footmisc}

\usepackage{url}        

\makeindex             

\def\la{\;
\raise0.3ex\hbox{$<$\kern-0.75em\raise-1.1ex\hbox{$\sim$}}\; }
\def\ga{\;
\raise0.3ex\hbox{$>$\kern-0.75em\raise-1.1ex\hbox{$\sim$}}\; }

\newcommand{\kms}{km~s$^{-1}\,$}
\newcommand{\ms}{m~s$^{-1}\,$}

\newcommand{\cmm}{cm$^{-3}\,$}
\newcommand{\etal}{{et al.}}
\newcommand{\DV}{$\Delta V$}
\newcommand{\hhho}{{H$_3$O$^+$}}
\newcommand{\nhhh}{{NH$_3$}}
\newcommand{\nnhp}{N$_2$H$^+$}
\newcommand{\nndp}{N$_2$D$^+$}
\newcommand{\hcccn}{HC$_3$N}

\begin{document}

\title*{Searching for Chameleon-like Scalar Fields}
\author{Sergei A. Levshakov, Paolo Molaro, Mikhail G. Kozlov, Alexander V. Lapinov, 
Christian Henkel, Dieter Reimers, Takeshi Sakai and Irina I. Agafonova}
\authorrunning{Levshakov, S. A., Molaro, P., Kozlov, M. G.,  et al.} 
\institute{S. A. Levshakov \at 
Physical-Technical Institute,
Polytekhnicheskaya Str. 26, 194021 St.~Petersburg, Russia, \email{lev@astro.ioffe.rssu.ru}
\and
P. Molaro \at INAF-Osservatorio Astronomico di Trieste, Via G. B. Tiepolo 11,
34143 Trieste, Italy, \email{molaro@oats.inaf.it}
\and
M. G. Kozlov \at Petersburg Nuclear Physics Institute, 188300 Gatchina, 
Russia, \email{mgk@mf1309.spb.edu}
\and
A. V. Lapinov \at Institute for Applied Physics, Uljanov Str. 46, 603950 Nizhny Novgorod, Russia,
\email{lapinov@appl.sci-nnov.ru}
\and
C. Henkel \at Max-Planck-Institut f\"ur Radioastronomie, Auf dem H\"ugel 69, D-53121 Bonn, Germany,
\email{p220hen@mpifr-bonn.mpg.de}
\and
D. Reimers \at Hamburger Sternwarte, Universit\"at Hamburg,
Gojenbergsweg 112, D-21029 Hamburg, Germany, \email{st2e101@hs.uni-hamburg.de}
\and
T. Sakai \at Institute of Astronomy, The University of Tokyo, Osawa, Mitaka, Tokyo
181-0015, Japan, \email{sakai@ioa.s.u-tokyo.ac.jp}
\and
I. I. Agafonova \at Physical-Technical Institute,
Polytekhnicheskaya Str. 26, 194021 St.~Petersburg, Russia, \email{ira@astro.ioffe.rssu.ru}
}
\maketitle

\abstract{
Using the 32-m Medicina, 45-m Nobeyama, and 100-m Effelsberg telescopes we found a statistically
significant velocity offset $\Delta V \approx 27 \pm 3$ \ms\ $(1\sigma)$
between the inversion transition in \nhhh (1,1) 
and low-$J$ rotational transitions in \nnhp (1-0) and \hcccn (2-1) arising in cold and dense
molecular cores in the Milky Way.
Systematic shifts of the line centers
caused by turbulent motions and velocity gradients,
possible non-thermal hyperfine structure populations, pressure and optical
depth effects are shown to be lower than or about 1 \ms\ and thus 
can be neglected in the total error budget.
The reproducibility of $\Delta V$ at the same facility (Effelsberg telescope) on a
year-to-year basis is found to be very good.
Since the frequencies of the inversion and
rotational transitions have different sensitivities to variations in 
$\mu \equiv m_{\rm e}/m_{\rm p}$,
the revealed non-zero $\Delta V$ may imply that $\mu$ changes 
when measured at high (terrestrial) and low (interstellar) matter densities as predicted by
chameleon-like scalar field models~-- candidates to the dark energy carrier. 
Thus we are
testing whether scalar field models have chameleon-type interactions with ordinary matter.
The measured velocity offset corresponds to the ratio
$\Delta \mu/\mu \equiv (\mu_{\rm space} - \mu_{\rm lab})/\mu_{\rm lab}$ of
$(26 \pm 3)\times10^{-9}$ ($1\sigma$).
}

\section{Introduction}
\label{sec:1}
This contribution sums up our results of differential measurements 
of the electron-to-proton mass ratio,
$\mu = m_{\rm e}/m_{\rm p}$,
carried out at 
the 32-m Medicina, 45-m Nobeyama, and 100-m Effelsberg telescopes
\cite{LMK08, MLK09, L10a, L10b}.
With high spectral resolution (FWHM $\sim$30-40 \ms),
we observed narrow emission lines (FWHM $< 200$ \ms) of N-bearing molecules arising in
cold and dense molecular cores 
of starless molecular clouds located in the Milky Way.
The cores are characterized by low kinetic temperatures  $T_{\rm kin} \sim 10$ K,
gas densities $n \sim 10^4 - 10^5$ \cmm, magnetic fields $B < 10$ $\mu$G, and
ionization degrees $x_e \sim 10^{-9}$ \cite{Di07}. 
The objective of this study is to probe the value of
$\Delta \mu/\mu \equiv (\mu_{\rm space} - \mu_{\rm lab})/\mu_{\rm lab}$
which is
predicted to be variable when measured at high (laboratory) and low (space) matter
density environments \cite{OP08}. 

A hypothetical variability of $\mu$ is thought to be due to
the scalar fields~-- candidates to the dark energy carrier,~--
which are ultra-light in cosmic vacuum but possess
an effectively large mass locally when they are coupled to
ordinary matter by the so-called chameleon mechanism \cite{KW04}.
Several possibilities to detect chameleons were discussed in \cite{Bur09,Dav09}.
First laboratory experiments constraining these models have been recently carried out
at Fermilab \cite{U10} and in the Lawrence Livermore National Laboratory \cite{Ry10}.

A subclass of chameleon models considers the couplings of a scalar
field to matter much stronger than gravitational and predicts that
fundamental physical quantities such as elementary particle masses
may depend on the local matter density, $\rho$, \cite{OP08}.
This means that a nonzero value of
$\Delta \mu/\mu$ is to be expected for all interstellar clouds irrespective of their position
and local matter density because the difference $\Delta \rho$
between the terrestrial environment in laboratory measurements
and dense interstellar molecular clouds is extremely large,
$\rho_\oplus/\rho_{\rm cloud} > 10^{10}$.

In the standard model (SM) of particle physics
the dimensionless mass ratio $\mu = m_{\rm e}/m_{\rm p}$ defines 
the ratio of the electroweak scale to the strong scale
since the mass of the electron $m_{\rm e}$ is proportional to the
Higgs vacuum expectation value and the mass of
the proton $m_{\rm p}$ is proportional to the quantum chromodynamics
scale $\Lambda_{\rm QCD}$ \cite{CFK09}.
The SM is extremely successful in explaining laboratory physics, 
but it has serious problems in astrophysics
where it completely fails to explain dark matter and dark energy. There are
many extensions of the SM including supersymmetry and different multidimensional theories which
introduce new particles as possible candidates for the dark matter and additional scalar fields
to describe the nature of dark energy.

A concept of dark energy with negative pressure appeared in physics
long before the discovery of the accelerating universe
through observations of nearby and distant Type Ia Supernovae \cite{Per98, Ri98}.
Early examples of dark energy in the form of a scalar field with a self-interaction potential
can be found in reviews \cite{Ca98} and \cite{PR03}.
If masses of the elementary particles are affected by scalar fields,
one can probe the dimensionless constant $\mu$
at different physical conditions by means of high precision spectral observations.
Namely, since the inversion and rotational molecular
transitions have different sensitivities to variations in $\mu$ \cite{FK07},
a nonzero $\Delta \mu/\mu$ causes an offset between
the radial velocities ($\Delta V \equiv V_{\rm rot} - V_{\rm inv}$)
of \textit{co-spatially} distributed molecules,
which, in turn, provide a measure of $\Delta \mu/\mu$.
When the inversion line belongs to \nhhh
we will herein call this procedure the ammonia method.

\section{The ammonia method}
\label{sec:2}
\nhhh\ is a molecule whose inversion frequencies are very sensitive to 
changes of $\mu$ because of the quantum mechanical tunneling of the
N atom through the plane of the H atoms. 
The inversion vibrational mode of \nhhh\ is described by a double-well
potential, the first two vibrational levels lying below the barrier.
The quantum mechanical tunneling splits these two levels into inversion doublets
providing a transition frequency that falls into the microwave range \cite{HoT83}.

The sensitivity coefficient
to $\mu$-variation of the \nhhh\ $(J,K) = (1,1)$ inversion
transition at 24 GHz is $Q_{\rm inv}=4.46$ \cite{FK07}. This means that
the inversion frequency scales as 
\begin{equation}
(\Delta \omega/\omega)_{\rm inv} \equiv (\tilde{\omega} - \omega)/\omega = 4.46(\Delta \mu/\mu)\ , 
\label{eq1}
\end{equation}
where $\omega$ and $\tilde{\omega}$ are the frequencies corresponding to the
laboratory value of $\mu$ and to an altered  $\mu$ in a low-density
environment, respectively.
In other words, the inversion transition sensitivity to $\mu$-variation is 4.46
times higher than that of molecular rotational
transitions, where $Q_{\rm rot}=1$ and thus, 
\begin{equation}
(\Delta \omega/\omega)_{\rm rot} = \Delta \mu/\mu \ .
\label{eq2}
\end{equation}

In astronomical spectra, any frequency shift $\Delta \omega$
is related to the radial velocity shift $\Delta V_r$
($V_r$ is the line-of-sight projection of the velocity vector)
\begin{equation}
\Delta V_r/c \equiv (V_r - V_0)/c
= (\omega_{\rm lab} - \omega_{\rm obs})/\omega_{\rm lab}\ ,
\label{eq3}
\end{equation}
where $V_0$ is the reference radial velocity, $c$ is the speed of light and
$\omega_{\rm lab}$, $\omega_{\rm obs}$ are
the laboratory and observed frequencies, respectively. Therefore, by comparing the
apparent radial velocity $V_{\rm inv}$ for the \nhhh\ inversion transition
with the apparent radial velocity  $V_{\rm rot}$ for rotational lines
originating in the same molecular cloud and moving with
radial velocity $V_0$ with respect to the local
standard of rest, one can
find from Eqs.~(\ref{eq1} - \ref{eq3})
\begin{equation}
\Delta \mu/\mu = 0.289(V_{\rm rot} - V_{\rm inv})/c
\equiv 0.289\Delta V/c\ .
\label{eq4}
\end{equation}

The velocity offset $\Delta V$  in Eq.(\ref{eq4}) can be expressed as the
sum of two components
\begin{equation}
\Delta V = \Delta V_\mu + \Delta V_n\ ,
\label{eq5}
\end{equation}
where $\Delta V_\mu$ is the shift due to $\mu$-variation and $\Delta V_n$
is a random component caused by the inhomogeneous distribution of molecules,
turbulent motions and velocity gradients,
possible non-thermal hyperfine structure populations, pressure and optical
depth effects, and instrumental imperfections (the so-called \textit{Doppler noise}).

We will assume that the Doppler noise component has a zero mean,
$\langle \Delta V_n \rangle = 0$ \kms, and a finite variance, $Var(\Delta V_n) < \infty$.
Then, the signal $\Delta V_\mu$ can be estimated statistically by
averaging over a data sample
\begin{equation}
\langle \Delta V \rangle = \langle \Delta V_\mu \rangle\ ,
\label{eq6}
\end{equation}
and
\begin{equation}
Var(\Delta V) = Var(\Delta V_\mu) + Var(\Delta V_n)\ .
\label{eq7}
\end{equation}
Equation (\ref{eq7}) shows that the measurability of the signal $\langle \Delta V_\mu \rangle$
depends critically on the value of the Doppler noise, $Var(\Delta V_n)$.
The Doppler noise 
can be reduced by the appropriate choice of molecular lines and molecular clouds.

To have similar Doppler velocity shifts the chosen molecular transitions should share as far as possible
the same volume elements.
\nhhh\ inversion transitions are usually detected
in dense molecular cores ($n \sim 10^4-10^5$ \cmm). 
Mapping of these cores in different molecular lines shows that
there is a good correlation between the NH$_3$,
\nnhp, and \hcccn\ distributions \cite{FM93, Htz04, Taf04, Pag09}. 
However, in some clouds \nhhh\ is not traced by \hcccn.
The most striking case is the
dark cloud TMC-1, where peaks of line emissions are offset
by 7 arcmin \cite{Ol88}.
Such a chemical differentiation together with velocity gradients within molecular cores
is the main source of the unavoidable Doppler noise in Eq.(\ref{eq5}).
Additional scatter in the $\Delta V_n$ values
may be caused by the
different optical depths of the hyperfine structure transitions
and by external electric and magnetic fields discussed below.

\section{Selection of targets}
\label{sec:3}
To minimize the Doppler noise component in Eq.(\ref{eq5}), a preliminary selection of molecular cores 
suitable for the most precise measurements  
of the velocity offsets \DV\ between rotational and inversion transitions
is required.
In general, molecular cores are
not ideal spheres and when observed at high angular resolutions
they frequently exhibit complex substructures.
The line profiles may be asymmetric because of non-thermal bulk motions.
Taking this into account, we formulate the following selection criteria:
\begin{enumerate}
\item[1.]
The line profiles are symmetric and well described by a single-component Gaussian model.
This selection increases the accuracy of the line center measurement.
Multiple line components may shift the barycenter and affect the velocity
difference between molecular transitions because, e.g., the ratio \nhhh/\hcccn\ may vary
from one component to another.
\item[2.]
The line widths do not exceed greatly the Doppler width caused by the thermal motion of material,
i.e., the non-thermal components (infall, outflow, tidal flow, turbulence) do not dominate
the line broadening. This ensures that selected molecular lines correspond to the same
kinetic temperature and arise co-spatially.
For the molecules in question we require the ratio of the Doppler $b$-parameters to be
$\beta = b$(\nhhh)$/b$(\hcccn) or $\beta = b$(\nhhh)$/b$(\nnhp), be $\beta \geq 1$.
\item[3.]
The spectral lines are sufficiently narrow ($b \sim 0.1-0.2$ \kms)
for the hyperfine structure (hfs) components to be resolved.
This allows us to validate
the measured radial velocity by means of different hfs lines of the same molecular transition.
\item[4.]
The spectral lines are not heavily saturated and their profiles are not
affected by optical depth effects.
The total optical depth of the \nhhh\ hf transitions is $\tau \leq 10$,
i.e., the optical depth of the strongest hf component is $\la 1$.
\end{enumerate}

The kinetic temperature $T_{\rm kin}$, and the nonthermal (turbulent)
velocity dispersion, $\sigma_{\rm turb}$, can be estimated from the line broadening Doppler
parameters $b = \sqrt{2}\sigma_v$, of, e.g., the \nhhh\ (1,1) and \hcccn\ (2--1) lines.
Here $\sigma_v$ is the line of sight velocity dispersion of the molecular gas within
a given cloud. If the two molecular transitions trace the same material 
and have the same nonthermal
velocity component, $\sigma_v$ is
the quadrature sum of the thermal $\sigma_{\rm th}$ and turbulent $\sigma_{\rm turb}$ 
velocity dispersions.
In this case a lighter molecule with a mass $m_l$ should have
a line width wider than with a heavier molecule with $m_h > m_l$.

For thermally dominated line widths ($\beta > 1$) and co-spatially distributed species
we obtain the following relations (e.g., \cite{FM93}):
\begin{equation}
T_{\rm kin} = m_l m_h (\sigma^2_l - \sigma^2_h) / k(m_h - m_l)\, ,
\label{eq8}
\end{equation}
and
\begin{equation}
\sigma^2_{\rm turb} = (m_h \sigma^2_h - m_l \sigma^2_l) / (m_h - m_l)\, ,
\label{eq9}
\end{equation}
where $k$ is Boltzmann's constant, and the thermal velocity dispersion 
$\sigma_{\rm th}$ is given by
\begin{equation}
\sigma_{{\rm th},i} = (k T_{\rm kin}/m_i)^{1/2}\, .
\label{eq10}
\end{equation}
It should be noted that
\hcccn\ is usually distributed in a volume of the molecular core larger than \nhhh\ since
N-bearing molecules trace the inner core, 
whereas C-bearing molecules occupy the outer part \cite{Di07}.
Such a differentiation may cause a larger nonthermal
component in the velocity distribution of \hcccn.
If both molecules are shielded from external incident radiation,
and the gas temperature mainly comes from heating by cosmic rays,
then a formal application of Eqs.(\ref{eq8}) and (\ref{eq9}) to the apparent line widths provides
a lower limit on $T_{\rm kin}$ and an upper limit on $\sigma_{\rm turb}$.
In molecular cores where 
the only source of heating are the cosmic rays and the cooling
comes from the line radiation mainly from  CO, a lower bound on the kinetic
temperature is about 8 K \cite{GL78}.
Thus, point 2 of the selection criteria requires $T_{\rm kin} \sim 8$-10~K if both
lighter and heavier molecules are distributed co-spatially.

\section{Preliminary results}
\label{sec:4}
In our preliminary single-pointing observations \cite{L10a} we studied 41 molecular cores
along 55 lines of sight. The NH$_3$(1,1), HC$_3$N(2-1), and
N$_2$H$^+$(1-0) transitions were observed with the 100-m Effelsberg, 32-m Medicina, and 45-m Nobeyama telescopes.
The analysis of the total sample revealed large systematic shifts and `heavy tails' of the \DV\
probability distribution function resulting in a poor concordance between three mean values:
the weighted mean 
$\langle \Delta V \rangle _{\scriptscriptstyle \rm W} = 27.4\pm4.4$ \ms\
(weights inversionally proportional to the variances), 
the robust redescending $M$-estimate 
$\langle \Delta V \rangle _{\scriptscriptstyle \rm M} = 14.1\pm4.0$ \ms,
and the median $\Delta V_{\rm med} = 17$ \ms.
The scatter in the \DV\ values reflects effects related to the gas kinematics
and the chemical segregation of one molecule with respect to the other (approximately 50\% of the targets
showed $\beta < 1$).

If we now consider molecular lines that fulfill the selection criteria, the sample size is
$n = 23$, i.e. two times smaller than the total data set.
Nevertheless, for this reduced sample we obtained better concordance:
$\langle \Delta V \rangle _{\scriptscriptstyle \rm W} = 20.7\pm3.0$ \ms,
$\langle \Delta V \rangle _{\scriptscriptstyle \rm M} = 21.5\pm2.8$ \ms, and  
$\Delta V_{\rm med} = 22$ \ms.

In this preliminary study we found 7 sources with thermally dominated motions which provide
$\langle \Delta V \rangle _{\scriptscriptstyle \rm W} = 21.1\pm1.3$ \ms, 
$\langle \Delta V \rangle _{\scriptscriptstyle \rm M} = 21.2\pm1.8$ \ms,
and $\Delta V_{\rm med} = 22$ \ms.

The most accurate result was obtained with the Effelsberg 100-m telescope:
$\Delta V = 27\pm4_{\rm stat}\pm3_{\rm sys}$ \ms\ (the value is slightly corrected in \cite{L10b}).
The systematic error in this estimate is due to uncertainties of the rest frequencies of the
HC$_3$N(2-1) hfs transitions since uncertainties of the \nhhh(1,1) hfs transitions are less than 1 \ms\
\cite{Ku67}.  
When interpreted in terms of the
electron-to-proton mass ratio variation, this gives
$\Delta\mu/\mu = (26\pm4_{\rm stat}\pm3_{\rm sys})\times10^{-9}$.

Thus, for the first time we obtained an astronomical spectroscopic estimate of the relative change in
the fundamental physical constant $\mu$ at the level of $10^{-9}$ which is $10^3$ times more
sensitive than the upper limits on $\Delta \mu/\mu$ 
obtained by the ammonia method from extragalactic observations \cite{FK07, Mur08, Men08, Hen09}.

\section{Mapping of cold molecular cores in \nhhh\ and \hcccn\ lines}
\label{sec:5}
In this section we report on new observations in which
we measure \DV\ at different positions
across individual clouds in order to test the reproducibility of the velocity offsets
in the presence of large-scale velocity gradients \cite{L10b}.
From the list of molecular cores observed with the Effelsberg 100-m telescope,
we selected several objects with symmetric profiles 
of the \nhhh(1,1) and \hcccn(2--1) hfs transitions. 
In these objects the line widths are thermally dominated,
i.e.,
the parameter $\beta = \sigma_v$(\nhhh)$/\sigma_v$(\hcccn) $\geq 1$.
The chosen targets are the molecular cores L1498, L1512, L1517B, and L1400K, which
have already been extensively studied in many molecular lines 
\cite{BM89, Le01, Le03, Taf04, Taf06, Cra05}.

The inversion line of ammonia \nhhh(1,1) at 23.694 GHz and the
rotation line of cyanoacetylene \hcccn(2--1) at 18.196 GHz
were measured with a K-band HEMT (high electron mobility transistor)
dual channel receiver,
yielding spectra with an angular resolution of HPBW~$\sim 40''$
in two orthogonally oriented linear polarizations. 
The measurements were carried out in frequency-switching mode
using a frequency throw of 5\,MHz. The backend was an FFTS
(Fast Fourier Transform Spectrometer) operated with its minimum
bandwidth of 20\,MHz providing simultaneously 16\,384 channels
for each polarization. The resulting channel separations are 15.4 \ms\
for \nhhh\ and 20.1 \ms\ for \hcccn. We note,
however, that the true velocity resolution is about two times
lower, FWHM~$\sim 30$ \ms\ and 40 \ms, respectively \cite{K06}. 
The sky frequencies were reset at the onset of each scan and the
Doppler tracking was used continuously to track Doppler shifts during the observations.
Observations started by measuring the continuum emission of
calibration sources and continued by
performing pointing measurements toward a source close to the spectroscopic
target.  The calibration is estimated
to be accurate to $\pm$15\% and the pointing accuracy to be higher than $10''$.

After corrections for the rounded frequencies,
the individual exposures were co-added to increase
the signal-to-noise ratio S/N.
The spectra were folded to remove the effects
of the frequency switch, and base lines were determined for each spectrum.
The resolved hfs components show no kinematic substructure and consist of an
apparently symmetric peak profile without broadened line wings or self-absorption
features. 
The line parameters, such as the total optical depth in the transition $\tau$
(i.e., the peak optical depth if all hyperfine components were placed at the same velocity),
the radial velocity $V_{lsr}$,
the line broadening Doppler parameter $b$, and the amplitude $A$, were obtained through fitting
of the \textit{one-component} Gaussian model to the observed spectra:
\begin{equation}
T(v) = A\cdot \left[ 1 - \exp(-t(v)) \right]\, ,
\label{eq11}
\end{equation}
with
\begin{equation}
t(v) = \tau\cdot \sum^k_{i=1}\, a_i\, \exp\left[ -{(v - v_i - V_{lsr})^2}/{b^2} \right]\, ,
\label{eq12}
\end{equation}
which transforms for optically thin transitions into
\begin{equation}
T(v) = A'\cdot \sum^k_{i=1}\, a_i\, \exp\left[ -{(v - v_i - V_{lsr})^2}/{b^2} \right]\, .
\label{eq13}
\end{equation}
The sum in (\ref{eq12}) and (\ref{eq13}) runs over the $k = 18$ and $k = 6$ hfs components of
the \nhhh(1,1) and \hcccn(2--1) transitions, respectively.

To test possible optical depth effects, 
we calculated two sets of the fitting
parameters: (1) based on the analysis of only optically thin satellite lines
with $\Delta F_1 \ne 0$,
and (2) obtained from the fit to the entirety of the \nhhh(1,1) spectrum
including the main transitions with $\Delta F_1 = 0$,
which have optical depths $\tau \approx 1$-2,
as can be inferred from the relative intensities of the hfs components.
The resulting $V_{lsr}$ values for all targets were within the 
$1\sigma$ uncertainty intervals.

To control another source of errors caused by instrumental imperfections,
we carried out \textit{repeated} observations at 10 offset positions:
two in L1498, L1517B, and L1400K, respectively, and four in L1512. 
The \DV\ dispersion resulting from these repeated measurements 
is $\sigma(\Delta V) = 2.0$ \ms.
This dispersion is lower than the calculated $1\sigma$ errors of the individual $\Delta V$ values,
which ensures that we are not missing any significant instrumental errors at the level of a few \ms.

We also checked the velocity offsets \DV\  obtained in our observations
with the 100-m Effelsberg telescope in Feb 2009 and Jan 2010.
The reproducibility of the velocity offsets at the same facility on a
year-to-year base is very good for L1498, L1512 and L1517B 
(concordance within 1$\sigma$ uncertainty intervals), 
except for L1400K where the central point is probably an outlier.

The comparison of the velocity dispersions
determined from the \nhhh(1,1) and \hcccn(2--1) lines
did not show any significant variations with position within each molecular core.
All data are consistent with thermally dominated line broadening.
In particular, the following weighted mean values were determined:
$\beta_{\scriptscriptstyle \rm L1498} = 1.24\pm0.03$, 
$\beta_{\scriptscriptstyle \rm L1512} = 1.32\pm0.05$,
$\beta_{\scriptscriptstyle \rm L1517B} = 1.11\pm0.01$, and 
$\beta_{\scriptscriptstyle \rm L1400K} = 1.23\pm0.04$.
The weighted mean values of the velocity dispersions for \nhhh\
range between
$\sigma_{\scriptscriptstyle \rm L1512} = 78\pm1$ \ms\ and
$\sigma_{\scriptscriptstyle \rm L1400K} = 86\pm1$ \ms,
and for \hcccn\ between
$\sigma_{\scriptscriptstyle \rm L1512} = 59\pm2$ \ms\ and
$\sigma_{\scriptscriptstyle \rm L1517B} = 74\pm1$ \ms.
This can be compared with the speed of sound
inside a thermally dominated region of a cold molecular core that is defined as
(e.g., \cite{Sh77})
\begin{equation}
v_s = (k T_{\rm kin}/m_0)^{1/2}\, ,
\label{eq14}
\end{equation}
where $m_0$ is the mean molecular mass.
With  $m_0 \approx 2.3$ amu for molecular clouds,
one has $v_s \approx 60\sqrt{T_{\rm kin}}$ \ms, which
shows that at the typical kinetic temperature of 10 K the nonthermal velocities
are in general subsonic, and that the selected targets do represent the quiescent material
at different distances from the core centers.

The gas temperature in the molecular cores was  
estimated from the apparent line widths using Eq.~(\ref{eq8}).
For L1498 we obtained a lower limit on the kinetic temperature
$T_{\rm kin} = 7.1\pm0.5$ K (average over 8 points), which is slightly lower
than $T_{\rm kin} = 10$ K measured by a different method from the
relative populations of the $(J,K) = (2,2)$ and (1,1) levels of \nhhh\
described by the rotational temperature $T^{21}_{\scriptscriptstyle \rm R}$
\cite{Taf04}.
For L1512 the temperature averaged over 11 points 
is $T_{\rm kin} = 9.6\pm0.6$ K, which is consistent with the value of 10 K.
The kinetic temperatures in the L1517B core in all measured positions
is well below 9.5 K~-- the value from \cite{Taf04}, 
which means that the nonthermal
velocity dispersions of \nhhh\ and \hcccn\ differ significantly and that both
species \textit{do not trace} the same material.
The three scanned points in the L1400K core
showed $T_{\rm kin} = 8.3\pm3.2$ K~-- close to the expected value of 10~K (measurements
of the gas temperature in this core were not performed in previous studies).

We found that, in general, the spatial fluctuations of $T_{\rm kin}$ do
not exceed a few kelvin implying uniform heating and
absence of the localized heat sources.
The kinetic temperature tends to rise with the distance
from the core center, which is in line with
the results from \cite{Taf04}. 

The radial velocity profiles along the different diagonal cuts toward the selected
targets are shown in Figs.~3 and 4 in \cite{L10b}.
The diagonal cut in L1498
exhibits coherently changing velocities of $V_{lsr}$(\hcccn) and $V_{lsr}$(\nhhh)
except one point at the core edge.
The velocity gradient is small,
$|\nabla V_{lsr}| \approx 0.5$ km~s$^{-1}$~pc$^{-1}$.
This picture coincides with the previously obtained results based on observations of
CO, CS, N$_2$H$^+$ and \nhhh\ in this core
and was interpreted as an inward flow \cite{Le01,Taf04}.
Taken together, all available observations classify L1498
as one of the most quiet molecular cores.
Thus, we can expect that the Doppler noise
(irregular random shifts in the radial velocities between different transitions)
is minimal in this core.

\begin{figure}[t]
\vspace{-5.0cm}
\includegraphics[scale=.7]{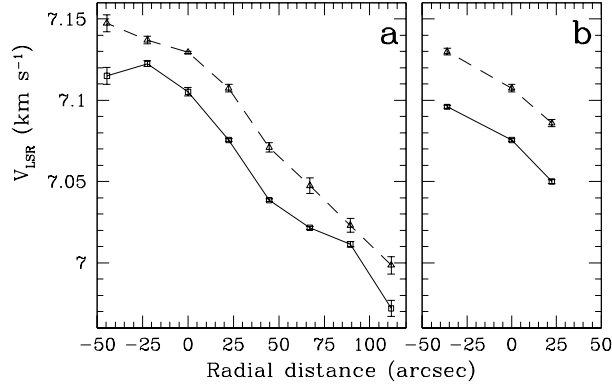}
\vspace{-4.0cm}
\caption
{
Example of the line-of-sight velocities of
\nhhh\ $(J,K) = (1,1)$ (squares) and \hcccn\ $J = 2-1$ (triangles)
at different radial distances along the main diagonal cut (panel {\bf a})
and in the perpendicular direction (panel {\bf b})
of the molecular core L1512. Shown are $1\sigma$ error bars.
For more detail, see \cite{L10b}.
}
\label{fg1}      
\end{figure}

In L1512, the $V_{lsr}$(\nhhh) and $V_{lsr}$(\hcccn) distributions are almost
parallel (Fig.~\ref{fg1}).
The same kinematic picture was obtained for this core in \cite{Le01} from observations of
CS and \nhhh\ lines and interpreted as a simple rotation around the center.
The velocity gradients derived from both \nhhh\ and \hcccn\ lines are similar,
$\nabla V_{lsr} \approx 1.5$ km~s$^{-1}$~pc$^{-1}$, and  consistent with the
gradient based on N$_2$H$^+$ measurements in \cite{Cas02}.
This means that \nhhh, \hcccn, and \nnhp\ trace the same gas, so the Doppler shifts \DV\
between them should be insignificant.

In the core L1517B, which is known to be very compact \cite{Le01}, we only observed 
the central  $30''\times30''$ where the velocities of \nhhh\ and \hcccn\ along 
two perpendicular cuts
do not change much:
$\nabla V_{lsr}$(\nhhh) $\approx 0.3$ km~s$^{-1}$~pc$^{-1}$,
$\nabla V_{lsr}$(\hcccn) $\approx 0.8$ km~s$^{-1}$~pc$^{-1}$,
and in the perpendicular direction
$\nabla V_{lsr}$(\nhhh) $\approx \nabla V_{lsr}$(\hcccn) $\approx -0.8$ km~s$^{-1}$~pc$^{-1}$.   
However, a wider area observation ($\approx 80''\times80''$)
of this core revealed an outward gas motion at the core periphery with
a higher velocity gradient,
$\nabla V_{lsr}$(N$_2$H$^+$) $\approx 1.1$ km~s$^{-1}$~pc$^{-1}$ \cite{Taf04},
which is consistent with earlier results on \nhhh\ observations \cite{Goo93}. 
This can lead to an additional shift in the radial velocity of \hcccn\ line since,
in general, the C-bearing molecules also trace a lower density gas 
component ($n < 10^4$ \cmm) in the envelope
of the molecular core. A higher nonthermal velocity dispersion of the \hcccn\ line than for \nhhh\
has already been mentioned above in regard to the temperature measurements in this core.

In L1400K we only observed three positions. Both molecules trace the same gradient of
$\nabla V_{lsr} \approx 1.9$ km~s$^{-1}$~pc$^{-1}$, which is in line with
$\nabla V_{lsr}$(N$_2$H$^+$) $= 1.8\pm0.1$ km~s$^{-1}$~pc$^{-1}$ derived in \cite{Cas02}.
The mapping in different molecular lines in \cite{Taf04} revealed that
L1400K deviates significantly from spherical symmetry and exhibits quite a complex
kinematic structure. In particular, the distributions of \nnhp\ and \nhhh\ do not coincide:
\nnhp\ has an additional component to the west of the center.
This explains why \cite{Cra05} reported 
a blue-ward skewness for this core,
$\theta = -0.42\pm0.10$, in the the N$_2$H$^+$ (1--0) hfs profiles,
whereas in our observations the \nhhh(1,1) hfs transitions are fully
symmetric: at the central position
the skewness of the resolved and single hfs 
component $F_1, F = 0, \frac{1}{2} \rightarrow 1, 1\frac{1}{2}$
of \nhhh\ is $\theta = -0.1\pm0.3$.

Our current measurements show very similar velocity shifts
$\langle \Delta V \rangle _{\scriptscriptstyle \rm M} = 25.8 \pm 1.7$ \ms\ 
and $28.0 \pm 1.8$ \ms\ ($M$-estimates)
for, respectively, the cores
L1498 and L1512 where the Doppler noise reaches minimal levels.
A higher shift $\langle \Delta V \rangle _{\scriptscriptstyle \rm M} = 46.9\pm3.3$ \ms\ 
is observed in the L1517B core~-- again in accord with its revealed kinematic structure, which
allows us to expect a higher radial velocity for the \hcccn\ line.
On the other hand, a lower value 
$\langle \Delta V \rangle _{\scriptscriptstyle \rm M} = 8.5\pm3.4$ \ms\ 
in L1400K may come from the irregular kinematic structure of
the core center, which could increase the radial velocity of the \nhhh\ line.

Thus, as reference velocity offset we chose the most robust $M$-estimate of the mean value from
the L1498 and L1512 cores:
$\langle \Delta V \rangle _{\scriptscriptstyle \rm M} = 26.9\pm1.2_{\rm stat}$ \ms.
Taking into account that the uncertainty of the \hcccn(2--1)
rest frequency is about 3 \ms, whereas that of \nhhh(1,1) is less than 1 \ms,
we finally have
$\langle \Delta V \rangle _{\scriptscriptstyle \rm M} = 26.9\pm1.2_{\rm stat}\pm3.0_{\rm sys}$ \ms.
When it is interpreted in terms of the electron-to-proton
mass ratio variation, as defined in Eq.(\ref{eq4}),
this velocity offset provides 
$\Delta \mu/\mu = (26\pm1_{\rm stat}\pm3_{\rm sys})\times10^{-9}$.

\section{Discussion and conclusions}
\label{sec:6}
In two molecular cores with the lowest Doppler noise L1498 and L1512, we
register very close values of the velocity offset \DV\ $\sim 27$ \ms\ between the
rotational transition \hcccn(2--1) and the inversion transition \nhhh(1,1).
These values coincide with the most accurate estimate obtained from
the Effelsberg dataset on 12 molecular clouds in the Milky Way \cite{L10a}.
Two other cores, L1517B and L1400K, exhibit velocity
shifts that are either higher ($\sim47$ \ms\ in L1517B) or lower ($\sim 9$ \ms\ in L1400K)
than the mean value, but the positive (L1517B) and
negative (L1400K) deflections from the mean can be explained from the observed kinematics
in these cores.

Of course, simultaneous observations of the
\nhhh(1,1) inverse transition  and rotational
transitions of some N-bearing molecules, such as  \nnhp(1-0) and \nndp(1-0) would
give a more accurate test. The main obstacle to this way is that the laboratory frequencies of
\nnhp\ and \nndp\ are known with accuracies not better than 14 \ms\ \cite{L10a}. Using
the \nnhp\ rest frequency from the Cologne Database for Molecular Spectroscopy
(CDMS) \cite{Mu05}
and observing with the Nobeyama 45-m telescope, we obtained a
velocity shift between \nnhp\ and \nhhh\  of $23.0\pm3.4$ \ms\ in L1498,  $24.5\pm4.3$ \ms\ in L1512,
and $21.0\pm5.1$ \ms\ in L1517B \cite{L10a}.
We note that similar shifts are indicated for the central parts of L1498
and L1517B in Fig.~11 in \cite{Taf04}, where these targets were observed with the 30-m IRAM telescope 
also using an
\nnhp\ rest frequency which is very close to the CDMS value.

Obviously, for more definite conclusions, new laboratory measurements of the
rest frequencies and new observations involving
other targets and other molecular transitions with different sensitivity coefficients $Q$ are required.
It has already been suggested to measure $\Lambda$-doublet lines of the light
diatomic molecules OH and CH \cite{Koz09}, microwave
inversion-rotational transitions in the partly deuterated ammonia
NH$_2$D and ND$_2$H \cite{KLL10},
low-laying rotational transitions in $^{13}$CO and 
the fine-structure transitions in C\,{\sc i}\ \cite{LMR10},
and tunneling and rotation
transitions in the hydronium ion H$_3$O$^+$\ \cite{KL10}. 
The fourth opportunity is of particular interest since the rest-frame frequencies of H$_3$O$^+$\
transitions are very sensitive to the variation of $\mu$,
and their sensitivity coefficients have \textit{different} signs.
For example, the
two lowest frequency transitions 
$J_K = 1_1^- \rightarrow 2_1^+$ and $J_K = 3_2^+ \rightarrow 2_2^-$ of
para-\hhho\ at, respectively, 307 and 364 GHz have $\Delta Q =
Q_{307} - Q_{364} = 14.7$, which is 4 times larger than $\Delta Q = 3.46$
from the ammonia method. This means that the offset $\Delta V \sim 27$ \ms,
detected with the ammonia method, should correspond to
a relative velocity shift between these transitions, $\Delta V =
V_{364} - V_{307}$, of about $100$ \ms\ if $\Delta \mu/\mu \approx 26\times10^{-9}$.
We consider the hydronium method \cite{KL10} as an important independent test of the  
$\Delta \mu/\mu$ value in the Milky Way.

To conclude, we note that in cold molecular cores with low
ionization degrees ($x_e \sim 10^{-8}-10^{-9}$) frequency shifts
caused by external
electric and magnetic fields and by the cosmic black body radiation-induced Stark effect
are less or about 1 m~s$^{-1}$ and cannot affect the revealed nonzero velocity offset
between the rotational and inversion transitions in the ammonia method. Detailed calculations of these
effects are given in \cite{L10b}.

Our current results tentatively support the hypothesis that the fundamental physical
constant~-- the electron-to-proton mass ratio~--
may differ in low-density environments from its terrestrial value.
This may be the consequence of the chameleon-like scalar field.
However, new laboratory measurements of the molecular rest frequencies and new observations involving other
molecular transitions and other targets 
are required to reach more definite conclusions.

\begin{acknowledgement}
We are grateful to the staffs of 
the Medicina 32-m, Nobeyama 45-m, and Effelsberg 100-m radio telescope
observatories for excellent support in our observations.
We thank Gabriella Schiulaz for her kind assistance in preparing the text.
The project has been supported in part by
DFG Sonderforschungsbereich SFB 676 Teilprojekt C4,
the RFBR grants No. 09-02-12223, 09-02-00352, and 08-02-92001,
the Federal Agency for Science and Innovations grant NSh-3769.2010.2,
the Program IV.12/2.5 of the Physical Department of the RAS.
\end{acknowledgement}

\end{document}